\documentclass{article}

\usepackage{amstext,amsmath,amssymb}
\usepackage[dvips]{color}
\usepackage{graphicx}
\usepackage{epsfig}
\usepackage{cite}
\usepackage{mcite}

\begin{document}

\setcounter{secnumdepth}{2}

\title{Definition of Systematic, Approximately Separable and Modular Internal Coordinates (SASMIC) for macromolecular simulation}

\author{Pablo Echenique$^{1,2}$ and J. L. Alonso$^{1,2}$
\vspace{0.4cm}\\ $^{1}$ {\small Departamento de F\'{\i}sica
Te\'orica, Facultad de Ciencias, Universidad de Zaragoza,}\\
{\small Pedro Cerbuna 12, 50009, Zaragoza, Spain.}\\
$^{2}$ {\small Instituto de Biocomputaci\'on y
F\'{\i}sica de los Sistemas Complejos (BIFI),}\\ {\small Edificio
Cervantes, Corona de Arag\'on 42, 50009, Zaragoza, Spain.}}

\date{\today}

\maketitle

\begin{abstract}

A set of rules is defined to systematically number the
groups and the atoms of organic molecules and, particularly, of polypeptides
in a modular manner. Supported by this numeration, a set of internal
coordinates is defined. These coordinates (termed Systematic, Approximately
Separable and Modular Internal Coordinates, SASMIC) are
straightforwardly written in Z-matrix form and may be directly
implemented in typical Quantum Chemistry packages. A number of Perl scripts
that automatically generate the Z-matrix files for polypeptides are
provided as supplementary material. The main difference
with other Z-matrix-like coordinates normally used in the literature
is that normal dihedral angles (``principal dihedrals'' in this work)
are only used to fix the orientation of whole groups and a somewhat
non-standard type of dihedrals, termed ``phase dihedrals'', are used
to describe the covalent structure inside the groups. This {\it
physical approach} allows to approximately separate soft and hard
movements of the molecule using only topological information and to
directly implement constraints. As an application, we use the
coordinates defined and ab initio quantum mechanical calculations to
assess the commonly assumed approximation of the free energy, obtained
from ``integrating out'' the side chain degree of freedom $\chi$, by
the Potential Energy Surface (PES) in the protected dipeptide
HCO-{\small L}-Ala-NH$_{2}$. We also present a sub-box of the Hessian
matrix in two different sets of coordinates to illustrate the
approximate separation of soft and hard movements when the coordinates
defined in this work are used.
\vspace{0.2cm}
\vspace{0.2cm}\\

\end{abstract}


\section{Introduction}
\label{sec:introduction}

The choice of the coordinates used to describe a molecule is an
important issue if computational considerations are to be taken into
account and the efficiency of the simulations is pursued. This choice
also affects the coding of applications. If cumbersomely defined
coordinates are used, an unnecessary complexity may be added to the
design of Monte Carlo movements, the construction and pruning of a
database of structures \cite{PE:Cha2002IJQC,PE:Sah2003JMS} or the
programming of molecular visualization and manipulation tools.

Suitable coordinates frequently used to describe arbitrary
conformations of molecules are the so-called ``internal'' or
``valence-type'' coordinates \cite{PE:Cra2002BOOK}. Their adequacy
stems from a number of characteristics: first, they are closely
related to chemically meaningful structural parameters, such as bond
lengths or bond angles; second, they are local, in the sense that each
one of them involves only a small number of (normally close) atoms in
its definition; and finally, there are only $3N-6$ of them (where $N$
is the number of atoms in the molecule), in such a way that the
overall rotation and translation have been naturally removed.

There also exists a family of coordinates
\cite{PE:Bak1996JCP,PE:vAr1999JCP,PE:Pai2000JCP,PE:Nem2004JCP},
extensively used in the inner calculations of many Quantum Chemistry
packages (such as Gaussian \cite{PE:Gaussian03} or GAMESS
\cite{PE:Sch1993JCC}) and based on the ``natural internal
coordinates'' originally proposed by Pulay and coworkers
\cite{PE:Pul1979JACS,*PE:Pul1992JCP,*PE:Fog1992JACS}, which are
defined through linear combinations of the original internals. These
coordinates are specially designed to describe normal-mode vibrations
in the immediate neighbourhood of energy minima and represent the best
choice for accelerating convergence of geometry optimizations in a
particular basin of attraction, via diagonal estimation of the Hessian
matrix \cite{PE:Nem2004JCP}. Accordingly, they maximally separate hard
and soft movements in these conditions. However, if the conformation
of the molecule is far from a minimum, this type of coordinates lose
great part of their meaning and they introduce many computational
difficulties without increasing the efficiency. Also, some of the
definitions are {\it redundant}
\cite{PE:Pul1979JACS,*PE:Pul1992JCP,*PE:Fog1992JACS,PE:Pai2000JCP},
i.e., they use a number of coordinates larger than the number of
degrees of freedom. In this work, we will only discuss coordinates,
such as internals or Cartesian, that may be conveniently used to
specify an {\it arbitrary} conformation of the system and that can be
directly related to simple geometrical variables.

In macromolecules, such as proteins, the number of degrees of freedom
is the main limiting factor when one tries to predict their behaviour
via computer modeling. Therefore, it is also advisable that the set of
coordinates chosen allows for a direct implementation of physically
meaningful constraints that reduce the dimensionality of the
conformational space considered. Most of the expressions used in
Statistical Mechanics or in Molecular Dynamics are best written in
Cartesian coordinates, however, the implementation of constraints
naturally appearing is far from being straightforward in these
coordinates. In internal coordinates, on the contrary, the approximate
separation of hard and soft movements of the system allows to easily
constrain the molecule
\cite{PE:Maz1989JBSD,*PE:Aba1989JBSD,*PE:Aba1994JCC} by setting the
hard coordinates (those that require a considerable amount of energy
to change noticeably) to constant values or to particular functions of
the soft coordinates. Moreover, in internal coordinates (and appealing
to some reasonable approximations), the Statistical Mechanics formulae
for the constrained system may be written in convenient closed form
\cite{PE:Go1976MM,PE:Kar1980MM}.

Still, although the bond lengths and bond angles are
customarily regarded as hard and their definition is
unproblematic, the same is not true for dihedral angles.
Some definitions of dihedrals may lead to difficulties or
to worse separation of hard and soft modes. Let us exemplify this
with a particular case:

Consider the definition of Z-matrix-like \cite{PE:Lev1999BOOK}
internal coordinates for the HCO-{\small L}-Ala-NH$_{2}$ molecule in
fig.~\ref{fig:num_ala}. Imagine that we ``position'' (i.e., we write
the corresponding Z-matrix row) every atom up to the hydrogen denoted
by H$_9$ and that we are now prepared to position the hydrogens in the
side chain (H$_{10}$, H$_{11}$ and H$_{12}$) via one bond length, one
bond angle and one dihedral for each one of
them\footnote{\label{foot:zmatrix_notation}We will denote by $(i,j)$
the bond length between atoms $i$ and $j$; by $(i,j,k)$, the bond
angle between the vectors ${\vec r}_{jk}$ and ${\vec r}_{ji}$; and by
$(i,j,k,l)$ the dihedral angle between the plane defined by the atoms
$i$, $j$ and $k$ and the one defined by $j$, $k$ and $l$.}. A choice
frequently seen in the literature
\cite{PE:Fre1992JACS,PE:Cha2002IJQC,PE:Sah2003JMS,PE:Jal1996CP,PE:Jal2002IJQC}
is the one shown in table~\ref{tab:bad_coord}.

\begin{table}[!ht]
\begin{center}
\begin{tabular}{cccc}
Atom name & Bond length & Bond angle & Dihedral angle \\
\hline \\[-8pt]
H$_{10}$ & (10,8) & (10,8,5) & $\gamma_1:=$(10,8,5,3) \\
H$_{11}$ & (11,8) & (11,8,5) & $\gamma_2:=$(11,8,5,3) \\
H$_{12}$ & (12,8) & (12,8,5) & $\gamma_3:=$(12,8,5,3)
\end{tabular}
\end{center}
\caption{\label{tab:bad_coord}{\small A part of the internal coordinates,
in Z-matrix form, of the protected dipeptide HCO-{\small L}-Ala-NH$_2$, as
frequently defined in the literature.}}
\end{table}

If we now perform the {\it gedanken experiment} that consists of
taking a typical conformation of the molecule and slightly moving each
internal coordinate at a time while keeping the rest constant, we find
that any one of the three dihedrals in the previous definition is a
hard coordinate, since moving one of them while keeping the other two
constant distorts the internal structure of the methyl group. Hence,
in these coordinates, the soft rotameric degree of freedom $\chi$,
which we know, for chemical arguments, that must
exist\footnote{\label{foot:rotation_barrier}According to our
calculations, at the RHF/6-31+G(d) level of the theory, the barrier
for crossing from one of the three equivalent minima to any of the
other two ranges from 3.1 to 6.8 kcal/mol, depending on the values of
the Ramachandran angles $\phi$ and $\psi$. Compare with the barriers
in $\phi$ or $\psi$ which may be as large as 20 kcal/mol depending on
the region of the Ramachandran map explored.}, is ill-represented. In
fact, it must be described as a {\it concerted} movement of the three
dihedrals. In references \cite{PE:Cha2002IJQC,PE:Sah2003JMS}, this
fact is recognized and the concept of ``related dihedrals'' is
introduced, however, no action is taken to change the definition of
the coordinates.

\begin{figure}[!ht]
\begin{center}
\epsfig{file=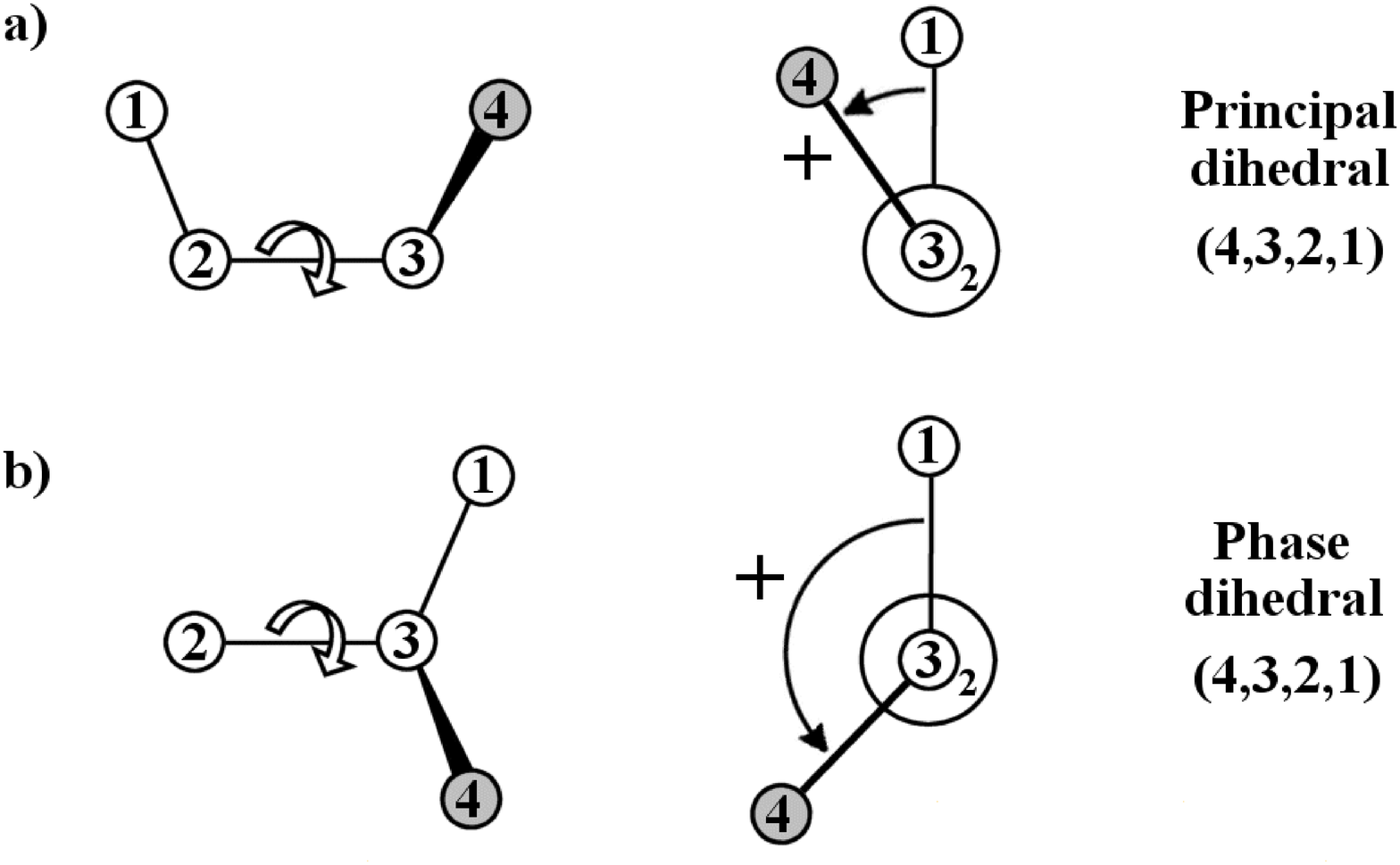,width=10cm}
\end{center}
\caption{\label{fig:dihedrals}{\small Two types of dihedral angles.
{\bf a)} {\it Principal dihedral}. Used to describe the rotation of whole
groups around bonds. {\bf b)} {\it Phase dihedral}. Used to describe
the internal covalent structure of groups. The positive sense of
rotation is indicated.}}
\end{figure}

In this work, using the ideas of R. Abagyan and coworkers
\cite{PE:Maz1989JBSD,*PE:Aba1989JBSD,*PE:Aba1994JCC}, we define a set
of rules to uniquely and systematically number the groups, the atoms
and define the internal coordinates of organic molecules and,
particularly, of polypeptides\footnote{\label{foot:IUPAC}IUPAC
conventions only define a numeration system for the groups, for the
branches and for some selected dihedral angles. They focus on
functional considerations and not in computational problems. For
related documents and references, see {\tt http://www.chem.qmul.ac.uk/
iupac/jcbn/}.}. The main difference with other Z-matrix-like
coordinates normally used in the literature
\cite{PE:Fre1992JACS,PE:Cha2002IJQC,PE:Sah2003JMS,PE:Jal1996CP,PE:Jal2002IJQC}
is that, instead of positioning each atom with a bond length, a bond
angle and a dihedral angle, we use normal dihedral angles (called,
from now on, ``principal dihedrals'') only to fix the orientation of
whole groups and a somewhat non-standard type of dihedrals, termed
``phase dihedrals'' by R. Abagyan and coworkers
\cite{PE:Maz1989JBSD,*PE:Aba1989JBSD,*PE:Aba1994JCC} (see
fig.~\ref{fig:dihedrals}), to describe the covalent structure inside a
group\footnote{\label{foot:two_angles}Another option may be to use, as
a third internal coordinate for each atom, another bond angle. This is
rather awkward, however, since two bond angles and a bond length do
not specify the position of a point in space. Any values of these
three coordinates (except for irrelevant degenerate cases) are
compatible with two different symmetrical positions and a fourth
number must be provided to break the ambiguity.}. This allows to
approximately separate soft and hard movements of the molecule using
only topological information (i.e., not knowing the exact form of the
potential) and to easily implement constraints by forcing the
coordinates that correspond to hard movements to take constant values
or ones that depend on the soft
coordinates\footnote{\label{foot:good_articles}In reference
\cite{PE:Yu2001JMS} they correctly take this approach into account
using out-of-plane angles instead of phase dihedrals, however, they do
not describe any rules for a general definition and their numeration
of the atoms is non-modular, as it proceeds first through the backbone
(see sec.~\ref{sec:definition_pro}).}.

In addition, the coordinates herein defined, are straightforwardly
cast into Z-matrix form and may be directly implemented in any Quantum
Chemistry package, such as Gaussian \cite{PE:Gaussian03} or GAMESS
\cite{PE:Sch1993JCC}. This is due to the fact that, although they
involve atoms whose covalent structure is different, the mathematical
construction of the two types of angles in fig.~\ref{fig:dihedrals} is
exactly the same, and the phase dihedrals are treated like principal
ones without any problem by the applications.

A number of Perl scripts are provided that number the atoms and
generate the coordinates herein defined for polypeptide chains. The
applications read a sequence file in which the different ionization
states of the titratable side chains, the tautomeric forms of
Histidine and several terminal groups may be specified. Then, an
output file is generated with the symbolic definition of the Z-matrix
of the molecule which may be directly pasted into the input files of
Gaussian \cite{PE:Gaussian03} or GAMESS \cite{PE:Sch1993JCC} (and,
upon slight modifications, of any Quantum Chemistry package that is
capable of reading Z-matrix format). These scripts may be found at
{\tt http://neptuno.unizar.es/files/public/gen\_sasmic/}.

Now, if we redo the example in table~\ref{tab:bad_coord} using phase
dihedrals, we must write the rows of the Z-matrix for the hydrogens in
the side chain as shown in table~\ref{tab:good_coord}.

\begin{table}[!ht]
\begin{center}
\begin{tabular}{cccc}
Atom name & Bond length & Bond angle & Dihedral angle \\
\hline \\[-8pt]
H$_{10}$ & (10,8) & (10,8,5) & $\chi:=${\bf (10,8,5,3)} \\
H$_{11}$ & (11,8) & (11,8,5) & $\alpha_1:=$(11,8,5,\underline{10}) \\
H$_{12}$ & (12,8) & (12,8,5) & $\alpha_2:=$(12,8,5,\underline{10})
\end{tabular}
\end{center}
\caption{\label{tab:good_coord}{\small A part of the internal
coordinates, in Z-matrix form, of the protected dipeptide
HCO-{\small L}-Ala-NH$_2$, as defined by the rules given in
secs.~\ref{sec:definition_gen} and \ref{sec:definition_pro}.}}
\end{table}

Where the angle {\bf (10,8,5,3)} is now the principal dihedral $\chi$
describing the relative rotation of the methyl group around the bond
length (8,5) and the other two are phase dihedrals that describe the
internal structure of the group and that are {\it pure} hard
coordinates (as far as can be told only from topological
information). However, one must point out that, although all bond
lengths, bond angles and phase dihedrals may be regarded as hard
coordinates, not all the principal dihedrals will be soft. Examples of
hard principal dihedrals are the ones that describe the rotation
around a double bond (or a triple one) or some of the principal
dihedrals in cyclic parts of molecules.

The {\it physical approach} described in this section, which should be
taken into account when designing internal coordinates, is embodied in
a set of rules for any organic molecule in
sec.~\ref{sec:definition_gen}, with a slightly different prescription
for polypeptide chains in sec.~\ref{sec:definition_pro}. The
systematic numeration introduced facilitates the computational
treatment of this type of systems and the rules given for polypeptide
chains ensure modularity \cite{PE:Cha2002IJQC,PE:Alo2004BOOK}, i.e.,
allows to add any residue with minimal modification of the already
existing notation and to easily construct databases of structures or
of Potential Energy Surfaces (PES).

The characteristics aforementioned have led us to term
the coordinates herein defined {\it Systematic, Approximately
Separable and Modular Internal Coordinates} (SASMIC).

In this work, we will only deal with the numeration of one isolated
molecule, however, the procedure described may be easily generalized
(and will be in future works) to systems of many molecules (an
important example being a macromolecular solute in a bath of solvent
molecules). This could be done using {\it ghost atoms} in a similar
manner to what is done in ref.~\cite{PE:Go1976MM}, to position the
center of mass of the system, and in
refs.~\cite{PE:Maz1989JBSD,*PE:Aba1989JBSD,*PE:Aba1994JCC}, to
actually define the coordinates of a system of molecules.

Finally, in sec.~\ref{sec:application}, we use the new coordinates and
ab initio quantum mechanical calculations in order to evaluate the
approximation of the free energy, obtained from ``integrating out''
the rotameric degree of freedom $\chi$, via the typical PES in the
protected dipeptide HCO-{\small L}-Ala-NH$_{2}$. This will be
relevant to design effective polypeptide potentials. We also
present a small part of the Hessian matrix in two different sets of
coordinates to illustrate the approximate separation of soft and hard
movements when the SASMIC defined in this work are
used. Sec.~\ref{sec:conclusions} is devoted to the conclusions.

\section{General numeration rules}
\label{sec:definition_gen}

\begin{figure}[!ht]
\begin{center}
\epsfig{file=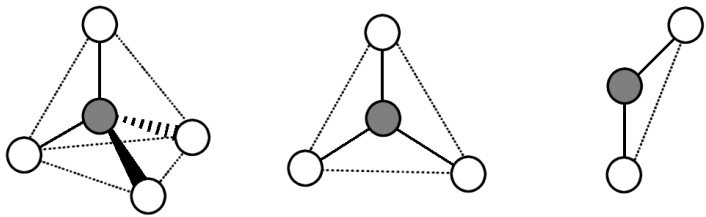,width=10cm}
\end{center}
\caption{\label{fig:groups}{\small Schematic representation of the
groups found in proteins.  From left to right: tetrahedral, triangular
and linear.}}
\end{figure}

First, we realize that any molecule may be formally divided in groups
such as those in fig.~\ref{fig:groups}. We will call ``centers'' the
shaded atoms in the figure and ``vertices'' the white ones. In
general, there may exist groups with more than four vertices, however,
in proteins, only groups with four or less vertices
occur\footnote{\label{foot:general}A fact that will not be used in the
definitions.}. Examples of tetrahedral groups are the one whose center
is the C$_{\alpha}$ in the backbone or the C$_{\beta}$ in the side
chain of alanine, triangular groups occur, for example, at the N or
the C' in the backbone, finally, linear groups may be found at the O
in the side chain of tyrosine or at the S in methionine (see
fig.~\ref{fig:20dipeptides}).

A particular atom may be vertex of different groups but may only be
center of one group. There exist atoms that are only vertices but
there do not exist atoms that are only centers, except in the case of
molecules with only one group. In the trivial case of diatomic
molecules (in which the only internal coordinate is a bond length),
neither of the previous definitions are possible, since we cannot
identify a group.

\begin{figure}[!ht]
\begin{center}
\epsfig{file=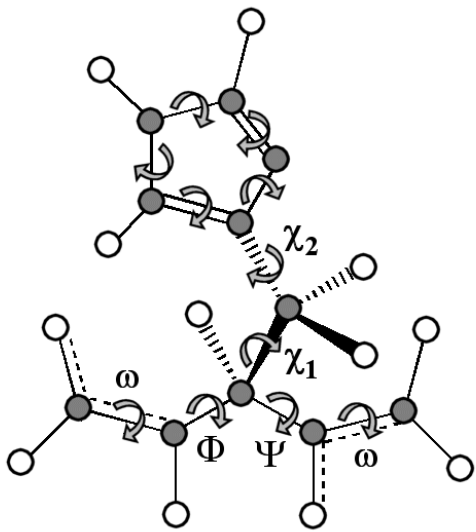,width=7cm}
\end{center}
\caption{\label{fig:int-ext_his}{\small Schematic representation of
the HCO-{\small L}-His-NH$_2$ model dipeptide (with the side chain
in its uncharged $\delta$ tautomeric form). Internal atoms are shown as
gray-filled circles, external ones as white-filled circles. Internal
bonds are indicated with curved arrows. Typical biochemical
definitions of some principal dihedrals are also shown.}}
\end{figure}

Atoms that are covalently bonded to more than one atom will be called
``internal atoms'' and are indicated as shaded circles in
fig.~\ref{fig:int-ext_his}. Atoms that are covalently attached to only
one internal atom will be called ``external atoms'' and are indicated
as white-filled circles in fig.~\ref{fig:int-ext_his}. In proteins,
only H and O may be external atoms.

In most macromolecular models (such as the Born-Oppenheimer
approximation used in sec.~\ref{sec:application}), nuclei are
considered point-like particles. Hence, rotation around bonds joining
external and internal atoms (termed ``external bonds'' or
``non-dihedral bonds'') is neglected, i.e., there are no internal
coordinates associated to this movement. On the other hand, rotation
around bonds joining two internal atoms (called ``internal bonds'' or
``dihedral bonds'' and indicated with curved arrows in
fig.~\ref{fig:int-ext_his}) is relevant and there may exist internal
coordinates describing it.

In order to conform with the {\it physical approach} stated in the
introduction, only one {\it golden rule} must be followed when
defining the internal coordinates:

\begin{quote}
{\it One principal dihedral, at most, must be defined on each internal bond.}
\end{quote}

The rest of the rules that will be given are mere
tidy conventions and systematics.

\subsection{General rules for numbering the groups}
\label{subsec:number_groups_gen}

First of all, we will divide the molecule in groups and number them.
To do this we proceed {\it by branches}, i.e., we choose the next
group following a linear sequence of covalently attached groups until
there is no possible next one, in which case, we either have finished
the numeration process or we start another branch. Every group is numbered one
time and it cannot be renumbered as the process continues. This
numeration is done for completeness and as a support for the numeration
of atoms and coordinates.

\begin{figure}[!ht]
\begin{center}
\epsfig{file=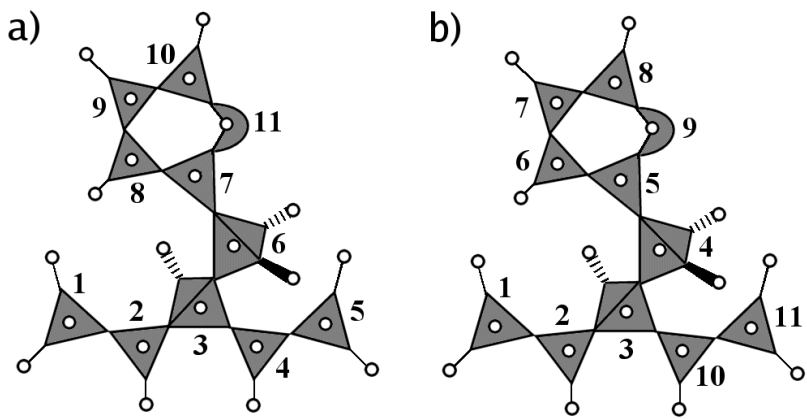,width=11cm}
\end{center}
\caption{\label{fig:groups_his}{\small Group identification and
numeration in the protected dipeptide HCO-{\small L}-His-NH$_2$ (with
the side chain in its uncharged $\delta$ tautomeric form). {\bf a)}
Following the general rules. {\bf b)} Following the special rules for
polypeptides. The different types of groups are shown as gray-filled
polyhedra.}}
\end{figure}

In fig.~\ref{fig:groups_his}a, we have implemented these {\it general}
rules in a protected histidine dipeptide. Later, in
sec.~\ref{sec:definition_pro}, while stating the rules for
polypeptides, we will remark the differences (which
will lead to fig.~\ref{fig:groups_his}b) and show the reasons for
the special prescriptions using the same example.

The rules are as follows:

\begin{enumerate}
\item[i)] The first group ($j=1$), is chosen, among those that are
 linked to the molecule via only one internal bond (termed ``terminal
 groups''), as {\it the one that has the greater
 mass}\footnote{\label{foot:group_mass}The mass of a group is defined
 as the sum of the atomic masses of its constituents.}. If two or
 more terminal groups have the same mass, we add the mass of their
 first neighbours to break the tie. If this does not lead to a
 decision, we proceed to the second neighbours and so on. If we run
 out of neighbours and there is still a tie, we choose a group
 arbitrarily among the ones that have been selected via this process
 and we indicate the convention. If there are no terminal groups, we
 perform this selection process among those groups that have {\it at least
 one external atom}\footnote{\label{foot:fullerene}The rare case in
 which there are neither terminal groups nor external atoms (such as
 C$_{60}$ fullerene) will not be treated here, although it would
 require only a small number of adjustments to the rules.}.
\item[ii)] If there is only one unnumbered group linked to group $j$,
 we number it as $j+1$, set $j=j+1$ and go to (ii).
\item[iii)] If there are two or more unnumbered groups linked to group
 $j$, we choose {\it the one with the greater mass} as in point (i), we
 number it as $j+1$, set $j=j+1$ and go to (ii).
\item[iv)] If there are no unnumbered groups linked to group $j$ but
 there are still unnumbered groups in the molecule, we set $j$ to {\it the
 number of the lowest numbered group that has unnumbered neighbours}
 (we prepare to start another branch) and we go to (ii).
\end{enumerate}

This process terminates when all the groups are numbered.

\subsection{General rules for numbering the atoms}
\label{subsec:number_atoms_gen}

\begin{figure}[!ht]
\begin{center}
\epsfig{file=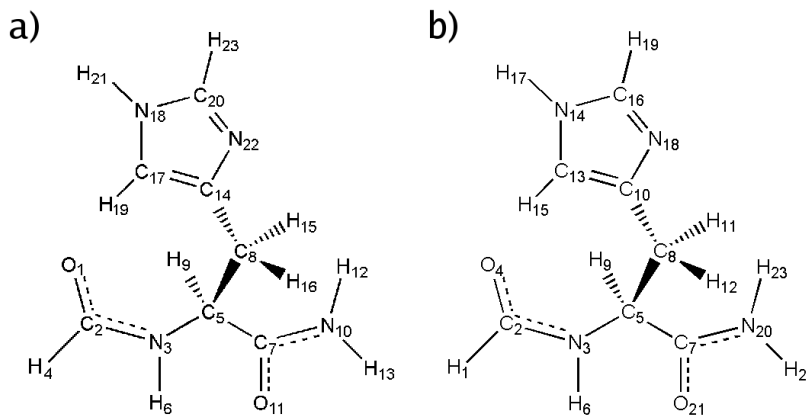,width=11cm}
\end{center}
\caption{\label{fig:num_his}{\small Atom numeration of the
protected dipeptide HCO-{\small L}-His-NH$_2$ (with the side chain
in its uncharged $\delta$ tautomeric form). {\bf a)} Following the general
rules. {\bf b)} Following the special rules for polypeptides.}}
\end{figure}

The atoms will be numbered in the order that they will be positioned
via internal coordinates in the Z-matrix.
As in the previous subsection, in fig.~\ref{fig:num_his},
these general rules, as well as the special rules for
polypeptides, are exemplified in a protected histidine dipeptide.
In sec.~\ref{sec:definition_pro}, we will remark the differences
and show the advantages of slightly modifying the prescription.

The rules are as follows:

\begin{enumerate}
\item[i)] The first atom ($k=1$), is chosen as {\it the heaviest of the
 external atoms in the first group}. If there are two or more
 candidates with the same mass, we choose arbitrarily and indicate
 the convention.
\item[ii)] The second atom ($k=2$) is the center of the first group and
 we set $j=1$ (the index of the group).
\item[iii)] If group $j+1$ exists and is covalently attached to group
 $j$, we number the unnumbered vertices of group $j$ {\it starting by
 the center of group $j+1$ and, then, in order of decreasing
 mass}. If, otherwise, group $j+1$ does not exist or it is not
 covalently attached to group $j$, {\it we simply number the
 unnumbered vertices of group $j$ in order of decreasing mass}. If, at
 any point, there are two or more candidates with the same mass, we
 choose arbitrarily and indicate the convention. \textsc{Exception}:
 If groups $j$ and $j+1$ belong to the same cyclic part of the
 molecule, the vertices of $j$ that are centers of groups (other than
 $j+1$) belonging to the same cycle {\it must not be numbered at this
 step} (for an example of this rule, see the numeration of C$_{17}$
 and N$_{22}$ in fig.~\ref{fig:num_his}a, or C$_{13}$ and N$_{18}$ in
 fig.~\ref{fig:num_his}b).
\item[iv)] If group $j+1$ does not exist, we have finished. Otherwise,
 we set $j=j+1$ and go back to (iii).
\end{enumerate}

\subsection{General rules for defining the internal coordinates}
\label{subsec:number_coor_gen}

Using the numeration for the atoms given in the previous section, we
give now a set of rules for defining the internal coordinates that
conform with the {\it physical approach} discussed in the introduction
of this work. The coordinates are written in Z-matrix form (see
table~\ref{tab:coor_his_gen}) for convenience and the rules are
applied to the protected dipeptide HCO-{\small L}-His-NH$_2$ (with the
side chain in its uncharged $\delta$ tautomeric form) using the
general numeration given in fig.~\ref{fig:num_his}a.

\begin{table}[!ht]
\begin{center}
\begin{tabular}{cccc}
Atom name & Bond length & Bond angle & Dihedral angle \\
\hline \\[-8pt]
O$_{1}$ & & & \\
C$_{2}$ & (2,1) & & \\
N$_{3}$ & (3,2) & (3,2,1) & \\
H$_{4}$ & (4,2) & (4,2,3) & (4,2,3,1) \\
C$_{5}$ & (5,3) & (5,3,2) & {\bf (5,3,2,1)} \\
H$_{6}$ & (6,3) & (6,3,2) & (6,3,2,5) \\
C$_{7}$ & (7,5) & (7,5,3) & {\bf (7,5,3,2)} \\
C$_{8}$ & (8,5) & (8,5,3) & (8,5,3,7) \\
H$_{9}$ & (9,5) & (9,5,3) & (9,5,3,7) \\
N$_{10}$ & (10,7) & (10,7,5) & {\bf (10,7,5,3)} \\
O$_{11}$ & (11,7) & (11,7,5) & (11,7,5,10) \\
H$_{12}$ & (12,10) & (12,10,7) & {\bf (12,10,7,5)} \\
H$_{13}$ & (13,10) & (13,10,7) & (13,10,7,12) \\
C$_{14}$ & (14,8) & (14,8,5) & {\bf (14,8,5,3)} \\
H$_{15}$ & (15,8) & (15,8,5) & (15,8,5,14) \\
H$_{16}$ & (16,8) & (16,8,5) & (16,8,5,14) \\
C$_{17}$ & (17,14) & (17,14,8) & {\bf (17,14,8,5)} \\
N$_{18}$ & (18,17) & (18,17,14) & {\bf (18,17,14,8)} \\
H$_{19}$ & (19,17) & (19,17,14) & (19,17,14,18) \\
C$_{20}$ & (20,18) & (20,18,17) & {\bf (20,18,17,14)} \\
H$_{21}$ & (21,18) & (21,18,17) & (21,18,17,20) \\
N$_{22}$ & (22,20) & (22,20,18) & {\bf (22,20,18,17)} \\
H$_{23}$ & (23,20) & (23,20,18) & (23,20,18,22)
\end{tabular}
\end{center}
\caption{\label{tab:coor_his_gen}{\small Internal coordinates in
Z-matrix form of the protected dipeptide HCO-{\small L}-His-NH$_2$
(with the side chain in its uncharged $\delta$ tautomeric form),
following the general rules. Principal dihedrals are indicated in bold
face.}}
\end{table}

The rules are as follows:

\begin{enumerate}
\item[i)] The positioning of {\it the first three atoms} is special.
 The corresponding rows of the Z-matrix are {\it always} as the
 ones in table~\ref{tab:coor_his_gen} (except, of course, for the
 chemical symbol in the first column, which may change).
\item[ii)] The positioning of the remaining vertices of group number
 1 (if there is any) is also special, their rows in the Z-matrix
 are:
\begin{center}
T$_i$ \ \ $(i,2)$ \ \ $(i,2,3)$ \ \ $(i,2,3,1)$
\end{center}
 Where T is the chemical symbol of the $i$-th atom, and $(i,2,3,1)$ is a
 phase dihedral.
\item[iii)] We set $i$ to {\it the number that follows that of the
 last vertex of the first group}.
\item[iv)] We choose $j$ as {\it the lowest numbered atom that is
 covalently linked to} $i$.
\item[v)] We choose $k$ as {\it the lowest numbered atom that is
 covalently linked to} $j$.
\item[vi)] If no principal dihedral has been defined on the bond
 $(j,k)$\footnote{\label{foot:golden}We say that a principal dihedral
 $(i,j,k,l)$ {\it is} ``on the bond $(j,k)$''.}, we choose $l$ as {\it
 the lowest numbered atom that is covalently linked to}
 $k$. Otherwise, we choose $l$ as {\it the second lowest numbered atom
 that is covalently linked to} $j$ (i.e., the lowest numbered atom
 that is covalently linked to $j$ and that is different from $k$, or,
 equivalently, the atom that was used to define the only principal
 dihedral on the bond $(j,k)$).
\item[vii)] The row of the Z-matrix that corresponds to atom $i$
 is:
\begin{center}
T$_i$ \ \ $(i,j)$ \ \ $(i,j,k)$ \ \ $(i,j,k,l)$
\end{center}
 Where T is the chemical symbol of atom $i$, $(i,j)$ is a bond
 length, $(i,j,k)$ is a bond angle and $(i,j,k,l)$ is a principal
 dihedral if the first case in point (vi) has occurred or
 a phase dihedral otherwise.
\item[viii)] If $i+1$ does not exist, we have finished. Otherwise,
 we set $i=i+1$ and go to (iv).
\end{enumerate}

\section{Special rules for polypeptides}
\label{sec:definition_pro}

The {\it numeration of the groups} in a polypeptide chain is the same as in
sec.~\ref{subsec:number_groups_gen} except for some details:

\begin{itemize}
\item Instead of using rule (i) for choosing the first group, we
 select\footnote{\label{foot:N-terminus}These three cases are
 the most frequent. If a different species is used to N-protect
 the polypeptide chain, a convention must be sought that
 also starts at the N-terminus.}:
 \begin{itemize}
 \item The {\it amino group} at the N-terminus (either charged or not)
  if the polypeptide is not N-protected.
 \item The {\it formyl group} at the N-terminus if the polypeptide
  is formyl-N-protected.
 \item The {\it methyl group} at the N-terminus if the polypeptide
  is acetyl-N-protected.
 \end{itemize}
 This is done because the primary structure of a polypeptide
 is normally presented from the N- to the C-terminus.
 In the case of HCO-{\small L}-His-NH$_2$ in fig.~\ref{fig:groups_his}, the
 agreement between rule (i) and this one is accidental.
\item When we must choose the next group to the one whose center
 is a C$_{\alpha}$ in the backbone, instead of applying rule (iii),
 which would yield the group at the C' as the next one
 (compare fig.~\ref{fig:groups_his}a and \ref{fig:groups_his}b),
 we choose {\it the first group in the side chain} (for residues
 that are different from glycine). Then, we resume the numeration
 as usual. This is done in order to ensure modularity, since,
 if we numbered following rule (iii), the backbone would be
 always numbered first and the whole numeration would have to
 be modified if we added a new residue to the chain.
\item Also for modularity reasons, we want to completely number
 the side chain before proceeding into the backbone. Hence, if we are
 numbering side chain atoms and the requirements to apply rule (iv)
 are fulfilled, instead of applying this rule, we set $j$ to {\it the
 number of the lowest numbered group that has unnumbered neighbours
 and that belongs to the side chain of the residue whose groups
 we are numbering.} Then, we go to (ii) as usual.
\end{itemize}

The {\it numeration of the atoms} in a polypeptide chain is the same as in
sec.~\ref{subsec:number_atoms_gen} except for some details:

\begin{itemize}
\item We seek that the principal dihedrals that are to be defined
 after numbering the atoms conform to the biochemical IUPAC conventions
 for the dihedrals $\phi$, $\psi$ and $\omega$ in the backbone.
 At the termini, we want that the atom where the C$_{\alpha}$ of
 the hypothetical residue $0$ or $N+1$ would occur is used to
 define the principal dihedrals. The general rules ensure this,
 except in two cases where a special convention must be given:
 \begin{itemize}
 \item If the polypeptide is formyl-N-protected, instead of
  applying rule (i) to choose the first atom, which would give
  the oxygen at the formyl group, we choose {\it the hydrogen
  at the formyl group}
  (compare fig.~\ref{fig:num_his}a and \ref{fig:num_his}b).
  Then, we resume the numeration at rule (ii).
 \item If the polypeptide is amide-C-protected, instead of
  applying rule (iii) and arbitrarily choosing one of the
  hydrogens in the terminal amide group before the other,
  we number {\it the trans hydrogen} before the other
  (compare fig.~\ref{fig:num_his}a and \ref{fig:num_his}b).
 \end{itemize}
\item Due to the special rules for the numeration of
 groups given above, the next group to the one at the C$_{\alpha}$
 is the first one in the side chain. If we applied rule (iii)
 for numbering the vertices of the C$_{\alpha}$-group,
 we would number first the center of the first group at the
 side chain and, then, the C' in the backbone. This
 would make the only principal dihedral defined on bond
 (C$_{\alpha}$,N) different from the conventional Ramachandran
 angle $\phi$. In order to avoid this, instead of applying
 rule (iii) at this point, we number the C' first among the
 unnumbered vertices of the C$_{\alpha}$-group and, then,
 resume the usual numeration process.
\end{itemize}

\begin{figure}[!ht]
\begin{center}
\epsfig{file=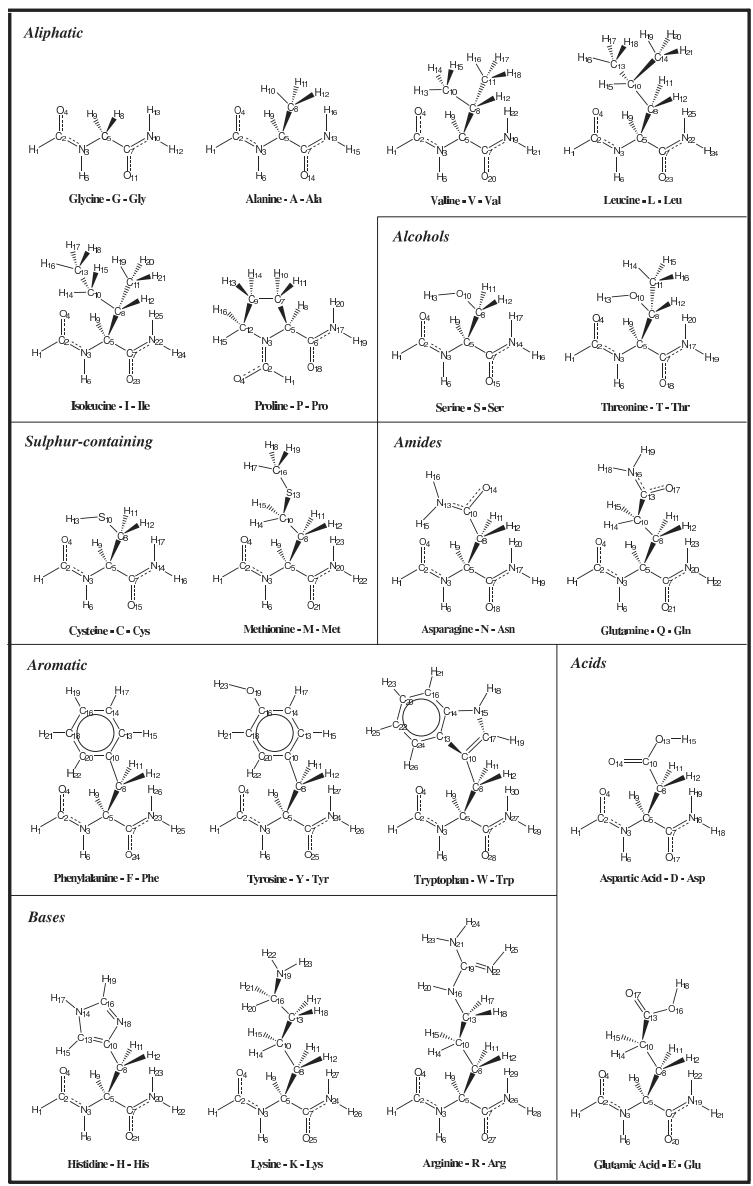,width=10cm}
\end{center}
\caption{\label{fig:20dipeptides}{\small Numeration of the left-handed
dipeptides HCO-{\small L}-X-NH$_2$, where X runs on the twenty
naturally ocurring amino acids (except for Glycine, which is the
achiral species HCO-Gly-NH$_2$). Uncharged side chains are displayed
and Histidine is shown in its $\delta$ tautomeric form. The rules used
for numbering are the special version for polypeptides.}}
\end{figure}

See fig.~\ref{fig:20dipeptides} for the numeration of the
twenty naturally ocurring amino acids with formyl-N-
and amide-C-protection.

Once the groups and the atoms have been numbered following these
special prescriptions, the {\it definition of the internal
coordinates} for polypeptide chains is identical to the one described
for the general case in sec.~\ref{subsec:number_coor_gen}.

\section{Application}
\label{sec:application}

\subsection{Theory}
\label{subsec:app_theory}

When a number of degrees of freedom are removed from the description
of the conformations of a physical system via their integrating
out in the partition function, the energy function that remains,
which describes the behaviour of the system only in terms of the rest
of the degrees of freedom, is a {\it free energy}. It depends on the
temperature and contains the entropy of the information that has been
averaged out as well as the enthalpy. However, it is frequent, when
studying the conformational preferences of model dipeptides in order
to use the information for designing effective potentials of
polypeptides
\cite{PE:Hal1996JCCa,*PE:Hal1996JCCb,*PE:Hal1996JCCc,*PE:Hal1996JCCd,*PE:Hal1996JCCe,PE:Jor1996JACS,PE:Cor1995JACS,*PE:Pon2003APC,PE:Bro1983JCC,*PE:Mac1998BOOK,PE:Mac2004JCC},
that the energy of these molecules be approximated by the Potential
Energy Surface (PES) in the bidimensional space spanned by the
Ramachandran angles $\phi$ and $\psi$
\cite{PE:Rod1998JMS,PE:Yu2001JMS,PE:Per2003JCC,PE:Mac2004JCC}. If we
recognize that the potential energy of the system in the
Born-Oppenheimer approximation (denoted by $V_{3N-6}$) depends on the
$3N-6$ internal coordinates, this surface (denoted by $V_2$) may be
defined as:

\begin{equation}
\label{eq:PES_def}
V_2(\phi,\psi):=\min_{Q^{\alpha}} \, V_{3N-6}(\phi,\psi,Q^{\alpha}) \ .
\end{equation}

Where $Q^{\alpha}$ denotes the rest of the internal coordinates.

The use of this surface, instead of a free energy function with the
$Q^{\alpha}$ degrees of freedom integrated out, is justified in the
approximation that these internal coordinates are {\it hard} and that
they are comparably much more difficult to excite at room temperature
than $\phi$ and $\psi$. If we assume that this is correct, these hard
degrees of freedom may be easily eliminated \cite{PE:Go1976MM} and the
partition function of the system may be written as follows:

\begin{equation}
\label{eq:Z_app}
Z=C \int\mathrm{d}\phi\,\mathrm{d}\psi\,\mathrm{d}Q^{\alpha}
 \,\mathrm{e}^{-\beta V_{3N-6}(\phi,\psi,Q^{\alpha})} \simeq
 C' \int\mathrm{d}\phi\,\mathrm{d}\psi
 \,\mathrm{e}^{-\beta V_2(\phi,\psi)} \ .
\end{equation}

Where $\beta:=1/RT$.

Note however that, in the ``flexible'' picture for the constraints,
this expression is correct only if we assume that the Jacobian
determinant of the change of coordinates from Cartesians to
$\{\phi,\psi,Q^{\alpha}\}$ and the determinant of the potential second
derivatives matrix with respect to the hard coordinates, both
evaluated at the equilibrium values, do not depend on $\phi$ and
$\psi$ (see ref.~\cite{PE:Go1976MM}). If, alternatively, we accept
the ``rigid'' picture for the constraints, we must ask that the
determinant of the induced metric tensor in the constrained
sub-manifold do not depend on $\phi$ and $\psi$ \cite{PE:Den2000MP}. If
these approximations (which will be reexamined in future works) do not
hold but the hardness of the $Q^{\alpha}$ degrees of freedom is still
assumed, the expressions in eq.~\ref{eq:Z_app} must be modified by
adding some correction terms to $V_2(\phi,\psi)$.

In eq.~\ref{eq:Z_app} for the partition function, one also may see
that, apart from the different multiplicative constants $C$ and $C'$,
which do not affect the expected values of observables, the use of the
PES $V_2(\phi,\psi)$ as the fundamental energy function of the system
is justified because it plays the same role as the whole potential
energy of the system in the first integral.

\begin{figure}[!ht]
\begin{center}
\epsfig{file=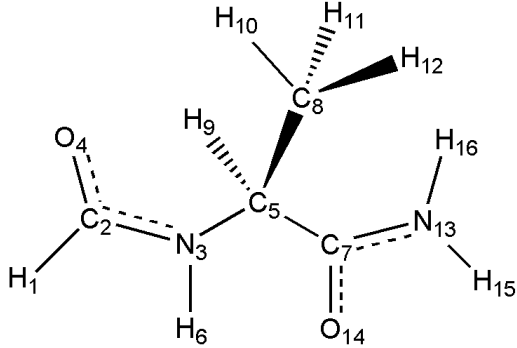,width=7cm}
\end{center}
\caption{\label{fig:num_ala}{\small Atom numeration of the protected
dipeptide HCO-{\small L}-Ala-NH$_2$ following the special rules for
polypeptides.}}
\end{figure}

However, although the hardness of the bond lengths, the bond angles
and even the dihedral $\omega$ in the peptide bond may be assumed,
this is not a good approximation for the rotameric degrees of freedom
in the side chains of residues. In the frequently studied
\cite{PE:Rod1998JMS,PE:Yu2001JMS} example of HCO-{\small L}-Ala-NH$_2$
(see fig.~\ref{fig:num_ala}), as it has already been said in
footnote~\ref{foot:rotation_barrier}, the side chain degree of freedom
$\chi$ must be regarded as {\it soft}. Still, although it is more
complex, a soft degree of freedom may also be averaged out if it
is considered convenient.

In this section, we will assume that the energy of the
formyl-alanine-amide dipeptide may be correctly approximated by a
Potential Energy Hypersurface (PEH) (denoted by $V_3$) that depends on
the Ramachandran angles $\phi$ and $\psi$ but also on the principal
dihedral $\chi$ that describes the rotation of the methyl group in the
side chain. Analogously to eq.~\ref{eq:PES_def}, its definition in
terms of the whole energy of the system is:

\begin{equation}
\label{eq:PEH_def}
V_3(\phi,\psi,\chi):=\min_{Q'^{\alpha}} \, 
 V_{3N-6}(\phi,\psi,\chi,Q'^{\alpha}) \ .
\end{equation}

Where $Q'^{\alpha}$ represents the internal coordinates that are not
$\phi$, $\psi$ or $\chi$.

Note, in addition, that the two definitions are related by the following
expression:

\begin{equation}
\label{eq:PES_PEH}
V_2(\phi,\psi)=\min_{\chi} \, V_3(\phi,\psi,\chi) \ .
\end{equation}

We will also assume for $V_3(\phi,\psi,\chi)$ the aforementioned
approximations that lead to eq.~\ref{eq:Z_app}, in such a way that we
can write (deliberately omitting the irrelevant multiplicative
constants):

\begin{eqnarray}
\label{eq:Z_app_hyper}
Z & \simeq & \int\mathrm{d}\phi\,\mathrm{d}\psi\,\mathrm{d}\chi
 \,\mathrm{e}^{-\beta V_3(\phi,\psi,\chi)} =
 \int\mathrm{d}\phi\,\mathrm{d}\psi\, Z(\phi,\psi) := \nonumber \\
 & := & \int\mathrm{d}\phi\,\mathrm{d}\psi\, 
 \mathrm{e}^{-\beta F(\phi,\psi)}\ .
\end{eqnarray}

Where we have defined:

\begin{equation}
\label{eq:F_def}
Z(\phi,\psi) := \mathrm{e}^{-\beta F(\phi,\psi)} := \int\mathrm{d}\chi\,
 \mathrm{e}^{-\beta V_3(\phi,\psi,\chi)} \ .
\end{equation}

This is what must be done in general when a soft degree of freedom is
needed to be integrated out in Statistical Mechanics
\cite{PE:Laz2003BPC} and the approximations in ref.~\cite{PE:Go1976MM}
cannot be made. The function $F(\phi,\psi)$ is a free energy because,
in general, it depends on the temperature and it contains the entropy
of the degree of freedom $\chi$ whose influence has been averaged.

We must remark at this point that, to integrate out the side chain
angle $\chi$ could be reasonable if one's aim is to use the ab initio
obtained information from a single dipeptide to include it in an
effective potential for simulating polypeptides. There is no point in
integrating out the Ramachandran angles $\phi$ and $\psi$, since the
conformation of the larger system will depend crucially on their
particular values, because they lie in the backbone of the molecule
and there are as many pairs $(\phi,\psi)$ as residues in the chain.
The side chain angle $\chi$, on the contrary, will
only influence its immediate surroundings and its importance could
be of different magnitude depending on the treatment that the
side chains are given in the model for the polypeptide.

In this context, if we wanted to use an energy function that does not
depend on $\chi$ (in some circumstances, a computational must), we
would have to perform the integral in the last term of
eq.~\ref{eq:F_def} and use $F(\phi,\psi)$ instead of $V_2(\phi,\psi)$,
since, as it has already been remarked, $\chi$ is not a hard
coordinate and the approximations needed to write eq.~\ref{eq:Z_app}
do not hold. Therefore, if we compare the last term in
eq.~\ref{eq:Z_app_hyper} with the last term in eq.~\ref{eq:Z_app}, we
see that, apart from additive constants that do not depend on $\phi$
and $\psi$ and that come from the multiplicative constants omitted,
the PES $V_2(\phi,\psi)$ must be understood as a candidate for {\it
approximating} the more realistic $F(\phi,\psi)$ and saving much
computational effort.

To assess the goodness of this approximation in the
particular case of formyl-alanine-amide is what will be done
in the following subsections.

\subsection{Methods}
\label{subsec:app_methods}

The ab initio quantum mechanical calculations have been done with the
package GAMESS \cite{PE:Sch1993JCC} under Linux. The coordinates used for the
HCO-{\small L}-Ala-NH$_2$ dipeptide in the GAMESS input files and the
ones used to ``move'' the molecule in the the automatic Perl scripts
that generated the input files are the SASMIC defined in
secs.~\ref{sec:definition_gen} and \ref{sec:definition_pro}.  They are
presented in table~\ref{tab:coor_ala_pro} indicating the name of the
conventional dihedral angles (see also fig.~\ref{fig:num_ala} for
reference). In the energy optimizations, on the contrary, they have
been converted to Delocalized Coordinates \cite{PE:Bak1996JCP} to accelerate
convergence.

\begin{table}[!ht]
\begin{center}
\begin{tabular}{cccc}
Atom name & Bond length & Bond angle & Dihedral angle \\
\hline \\[-8pt]
H$_{1}$ & & & \\
C$_{2}$ & (2,1) & & \\
N$_{3}$ & (3,2) & (3,2,1) & \\
O$_{4}$ & (4,2) & (4,2,3) & (4,2,3,1) \\
C$_{5}$ & (5,3) & (5,3,2) & $\omega_0:=${\bf (5,3,2,1)} \\
H$_{6}$ & (6,3) & (6,3,2) & (6,3,2,5) \\
C$_{7}$ & (7,5) & (7,5,3) & $\phi:=${\bf (7,5,3,2)} \\
C$_{8}$ & (8,5) & (8,5,3) & (8,5,3,7) \\
H$_{9}$ & (9,5) & (9,5,3) & (9,5,3,7) \\
H$_{10}$ & (10,8) & (10,8,5) & $\chi:=${\bf (10,8,5,3)} \\
H$_{11}$ & (11,8) & (11,8,5) & (11,8,5,10) \\
H$_{12}$ & (12,8) & (12,8,5) & (12,8,5,10) \\
N$_{13}$ & (13,7) & (13,7,5) & $\psi:=${\bf (13,7,5,3)} \\
O$_{14}$ & (14,7) & (14,7,5) & (14,7,5,13) \\
H$_{15}$ & (15,13) & (15,13,7) & $\omega_1:=${\bf (15,13,7,5)} \\
H$_{16}$ & (16,13) & (16,13,7) & (16,13,7,15)
\end{tabular}
\end{center}
\caption{\label{tab:coor_ala_pro}{\small Internal coordinates in
Z-matrix form of the protected dipeptide HCO-{\small L}-Ala-NH$_2$,
following the special rules for polypeptides. Principal dihedrals are
indicated in bold face and their typical biochemical name is given.}}
\end{table}

First, we have calculated the typical PES $V_2(\phi,\psi)$ defined in
eq.~\ref{eq:PES_def} in a regular 12x12 grid, with both $\phi$ and
$\psi$ ranging from $-165^{o}$ to $165^{o}$ in steps of $30^{o}$.
This has been done by running energy optimizations at the
\mbox{RHF/6-31+G(d)} level of the
theory\footnote{\label{foot:whythislevel}Since this is only an
exploratory study, the \mbox{RHF/6-31+G(d)} level of the theory is
considered enough to show the general trend and to illustrate the
procedure to be followed in a more exhaustive assessment with more
reliable ab initio methods.}, freezing the two Ramachandran angles at
each value on the grid, starting from geometries previously optimized
at a lower level of the theory and setting the gradient convergence
criterium to {\tt OPTTOL}=0.0001 and the self-consistent Hartree-Fock
convergence criterium to {\tt CONV}=0.00001.

Then, at each grid point, we have defined another one-dimensional grid
in the coordinate $\chi$ that ranges from $\chi_0(\phi,\psi)-50^{o}$
to $\chi_0(\phi,\psi)+60^{o}$ in steps of $10^{o}$, where
$\chi_0(\phi,\psi)$ is one of the three equivalent equilibrium values
(selected arbitrarily) of this degree of freedom at each point of the
original PES. This partition in 12 points spans one third of the
$\chi$-space, but it is enough for computing the integrals because the
surface $V_3(\phi,\psi,\chi)$ has exact three-fold symmetry in $\chi$
(note, for example, that the value of $V_3$ at
$\chi_0(\phi,\psi)-60^{o}$ would be equal to the one at
$\chi_0(\phi,\psi)+60^{o}$). Next, we have run energy optimizations,
with the same parameters described above and at the same level of
theory, at each point of the $\chi$-grid for every grid-value of the
PES (i.e., freezing the three angles). The starting geometries have
been automatically generated via Perl scripts taking the final
geometries in the $(\phi,\psi)$-grid and systematically changing
$\chi$. Note that this amounts to only changing the principal
dihedral (10,8,5,3) in the \mbox{Z-matrix} in table~\ref{tab:coor_ala_pro};
with poorly designed coordinates that did not separate the hard modes
from the soft ones, this process would have been
more difficult and rather unnatural.

After all the optimizations ($\sim$ 54 days of CPU time in 3.20 GHz
PIV machines), we have 12x12x12=1728 points with grid coordinates
\mbox{$(\phi_i,\psi_j,\chi_k)$} \mbox{$i,j,k=1 \ldots 12$} of the
function $V_3(\phi,\psi,\chi)$ and we may approximate the
integral defining $F(\phi,\psi)$ in eq.~\ref{eq:F_def} by a finite
sum:

\begin{eqnarray}
\label{eq:F_def_sum}
&& F(\phi_i,\psi_j) := -RT \, \ln{ \left ( \sum_k 
 \mathrm{e}^{-\beta V_3(\phi_k,\psi_j,\chi_k)} \right )} = \nonumber \\
&& = -RT \, \ln{ \left ( \sum_k 
 \mathrm{e}^{-\beta [V_3(\phi_k,\psi_j,\chi_k)-\langle V_3 \rangle
 (\phi_i,\psi_j)]} \right )} - \langle V_3 \rangle (\phi_i,\psi_j) \ .
\end{eqnarray}

Where additive constants arising from the three-fold symmetry
in the coordinate $\chi$ have been discarded.

The harmless quantity $\langle V_3 \rangle(\phi,\psi)$, defined as:

\begin{equation}
\label{eq:medV3_def}
\langle V_3 \rangle(\phi_i,\psi_j):=\frac{1}{12}
 \sum_kV_3(\phi_k,\psi_j,\chi_k) \ ,
\end{equation}

has been introduced in order for the values of the exponential
function to be in the precision range of the computer.

Analogously, the average energy may be computed via:

\begin{eqnarray}
\label{eq:U_def_sum}
&& U(\phi_i,\psi_j) := \frac{ \displaystyle \sum_k V_3(\phi_k,\psi_j,\chi_k)
 \mathrm{e}^{-\beta V_3(\phi_k,\psi_j,\chi_k)} }
 {\displaystyle \sum_k \mathrm{e}^{-\beta V_3(\phi_k,\psi_j,\chi_k)}} =
 \nonumber \\
&& = \frac{ \displaystyle \sum_k V_3(\phi_k,\psi_j,\chi_k)
 \mathrm{e}^{-\beta V_3(\phi_k,\psi_j,\chi_k)
 -\langle V_3 \rangle(\phi_i,\psi_j)]} }
 {\displaystyle \sum_k \mathrm{e}^{-\beta [V_3(\phi_k,\psi_j,\chi_k)
 -\langle V_3 \rangle(\phi_i,\psi_j)]}} = \ .
\end{eqnarray}

And, finally, we extract the entropy from:

\begin{equation}
\label{eq:FUmTS}
F(\phi_i,\psi_j)=U(\phi_i,\psi_j)-TS(\phi_i,\psi_j) \ .
\end{equation}

Additionally, apart from the calculations needed to integrate out
$\chi$, we have also performed an unconstrained geometry optimization
in the basin of attraction of the local minima of the PES normally
known as ${\gamma}_{\mathrm{L}}$ or C7$_{\mathrm{eq}}$ depending on
the author \cite{PE:Per2003JCC}. This calculation was done at the
MP2/6-31++G(d,p) level of the theory and with the same values of the
variables {\tt OPTTOL} and {\tt CONV} than the ones used in the PES
case. The starting geometry was the final structure corresponding to
the point $(-75^{o},75^{o})$ of the PES calculations at the lower
level of the theory described in the preceding paragraphs.

In the local minimum found, we have computed the Hessian
matrix (also at MP2/6-31++G(d,p)) in two different sets of coordinates:
the properly defined SASMIC shown in table~\ref{tab:coor_ala_pro}
and an ill-defined set in which the lines corresponding to the
hydrogens H$_{10}$, H$_{11}$ and H$_{12}$ in the side chain
have been substituted by those in table~\ref{tab:bad_coord}.
This is done to numerically illustrate the better separation
of the hard and soft modes achieved by the internal coordinates
defined in this work.

\subsection{Results}
\label{subsec:app_results}

\begin{figure}[!ht]
\begin{center}
\epsfig{file=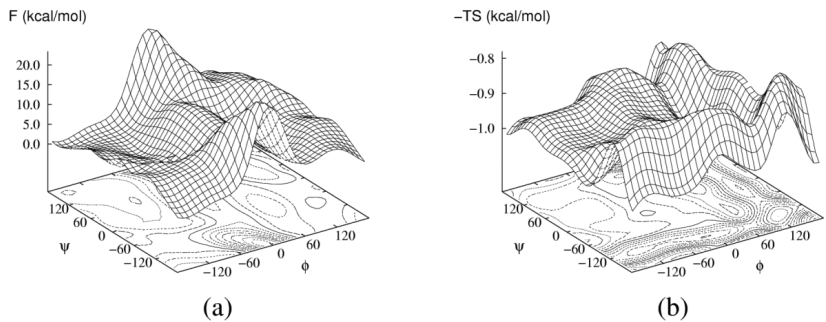,width=12cm}
\end{center}
\caption{\label{fig:rama_plots}{\small Ramachandran plots of {\bf (a)}
the free energy $F(\phi,\psi)$ and {\bf (b)} $-TS(\phi,\psi)$ in the
model dipeptide HCO-{\small L}-Ala-NH$_2$.}}
\end{figure}

In order to assess if $V_2(\phi,\psi)$ could be considered a good
approximation of $F(\phi,\psi)$, we have used a statistical quantity,
defined in \cite{PE:Alo2005JCC}, which measures the typical error that
one makes in the energy differences between arbitrary pairs of
conformations of the system if one effective potential is used instead
of the other. If we measure this {\it distance} between $F(\phi,\psi)$
and $V_2(\phi,\psi)$, using the 144 points in the $(\phi,\psi)$-grid,
we obtain:

\begin{equation}
\label{eq:d}
d(F,V_2)=0.098\ RT \ .
\end{equation}

We present the result in units of $RT$ (at $300^{o}$ K, where
$RT\simeq 0.6$ kcal/mol) because it has been argued in
\cite{PE:Alo2005JCC} that, if the distance between two different
approximations of the energy of the same system is less than $RT$, one
may safely substitute one by the other without altering the relevant
physical properties. {\it In this case, this criterium is widely
satisfied.} Moreover, if one assumes that the effective energy studied
will be used to construct a polypeptide potential and that the latter
will be designed as simply the sum of mono-residue ones (making each
term suitably depend on different pairs of Ramachandran angles), then,
the number $N_{res}$ of residues up to which one may go keeping the
distance between the two approximations of the the $N$-residue
potential below $RT$ is (see ref. \cite{PE:Alo2005JCC}):

\begin{equation}
\label{eq:Nres}
N_{res}(F,V_2)=\left ( \frac{RT}{d(F,V_2)} \right )^{2} \simeq 104 \ .
\end{equation}

The goodness of the approximation in this case is much due to
the simplicity and small size of the side chain of the
alanine residue and also to the fact that the dipeptide is
isolated. For bulkier residues included in polypeptides,
we expect the difference between $F(\phi,\psi)$ and $V_2(\phi,\psi)$
to be more important.

Although the essential result is the one stated in the previous
paragraphs, we wanted to look in more detail at the origin of the
differences between $F(\phi,\psi)$ and $V_2(\phi,\psi)$. For this,
we have first subtracted from $F(\phi,\psi)$, $U(\phi,\psi)$ and
$V_2(\phi,\psi)$ the same constant reference ($\min \,
F(\phi,\psi)$)\footnote{\label{foot:fundamental}At the level of the
theory used in the calculations, the minimum of $F(\phi,\psi)$ in the
grid is -414.7985507934 hartree.} in order to render the numerical
values more manageable and to minimize the statistical error of the
$y$-intercept in the linear fits \cite{PE:Bev2003BOOK,PE:Pre2002BOOK}
that will be made in the following.

\begin{figure}[!ht]
\begin{center}
\epsfig{file=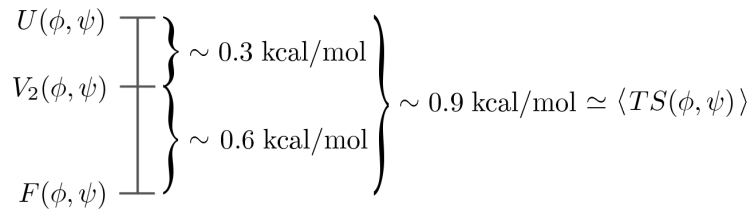,width=10cm}
\end{center}
\caption{\label{fig:offsets}{\small Relative offsets among
the thermodynamical surfaces involved in the study.}}
\end{figure}

Then, fitting $U(\phi,\psi)$ against $V_2(\phi,\psi)$, we have found that
they are more correlated than $F(\phi,\psi)$ and $V_2(\phi,\psi)$
(compare the Pearson's correlation coefficient, $r(U,V_2)=0.999999$
vs. $r(F,V_2)=0.999954$, and the aforementioned distance,
$d(U,V_2)=0.015\ RT$ vs. $d(F,V_2)=0.098\ RT$), and that they are
separated by an almost constant offset: $V_2(\phi,\psi)$ is $\sim 0.3$
kcal/mol lower that $U(\phi,\psi)$ (on the other hand,
$V_2(\phi,\psi)$ is $\sim 0.6$ kcal/mol higher than
$F(\phi,\psi)$). Hence, the three Ramachandran surfaces
$F(\phi,\psi)$, $U(\phi,\psi)$ and $V_2(\phi,\psi)$ are very similar,
except for an offset. In fig.~\ref{fig:rama_plots}a, $F(\phi,\psi)$ is
depicted graphically and, in fig~\ref{fig:offsets}, the relative
offsets among the three energies are schematically shown.

Contrarily, the entropy (we use $TS(\phi,\psi)$ in order to deal with
quantities that have units of energy), which may be found in
fig.~\ref{fig:rama_plots}b, and whose average magnitude is $\sim 0.9$
kcal/mol, is almost uncorrelated with $F(\phi,\psi)$, $U(\phi,\psi)$
and $V_2(\phi,\psi)$, being the correlation coefficients
$r(TS,F)=0.382$, $r(TS,U)=0.379$ and $r(TS,V_2)=0.381$,
respectively. Hence, given that $d(U,V_2)$ is almost an order of
magnitude lower than $d(F,V_2)$, it is reasonable to conclude that the
greatest part of the (little) noise between $F(\phi,\psi)$ and
$V_2(\phi,\psi)$ comes from the entropic term $-TS(\phi,\psi)$. This
is supported by the fact that the difference
$F(\phi,\psi)-V_2(\phi,\psi)$ is highly correlated with $TS(\phi,\psi)$,
being the correlation coefficient $r(F-V_2,TS)=0.998$.

Finally, and in order to illustrate the better separation of the hard
and soft modes achieved by the internal coordinates defined in this
work, we have calculated the Hessian matrix in the minimum
${\gamma}_{\mathrm{L}}$ (also C7$_{\mathrm{eq}}$) in two different
sets of coordinates. They are described at the end of
sec.~\ref{subsec:app_methods} and they correspond to the SASMIC set,
defined according to the rules given in sec.~\ref{sec:definition_pro},
and a set in which the coordinates that position the hydrogens in the
side chain have been ill-defined.

\begin{figure}[!ht]
\begin{center}
\epsfig{file=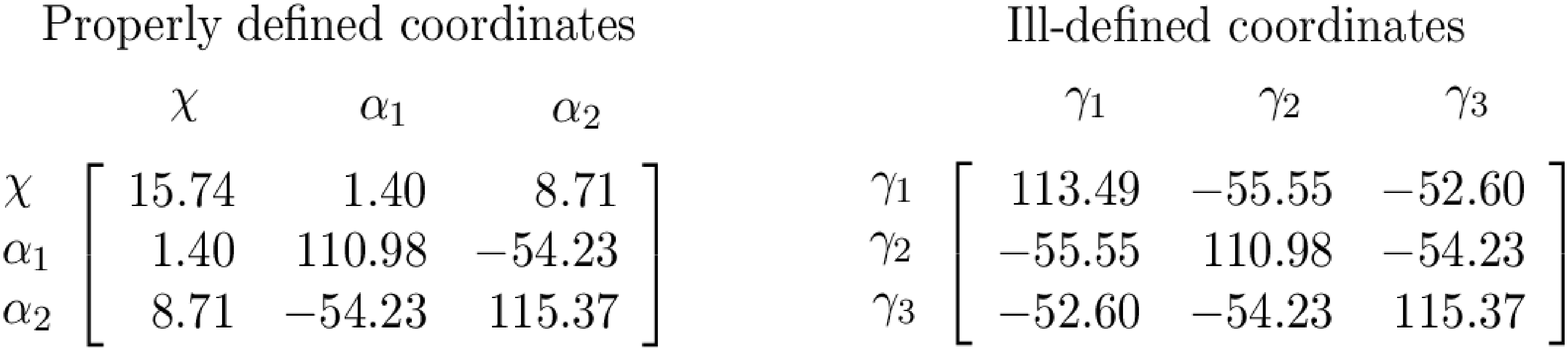,width=10cm}
\end{center}
\caption{\label{fig:hessians}{\small Sub-boxes of the Hessian matrix
in the minimum ${\gamma}_{\mathrm{L}}$ (also C7$_{\mathrm{eq}}$)
corresponding to the coordinates defined in
tables~\ref{tab:good_coord} and \ref{tab:bad_coord}.
The quantities are expressed in \mbox{kcal/mol $\cdot$ rad$^{-2}$}.
See the text for more details.}}
\end{figure}

In fig.~\ref{fig:hessians}, we present the sub-boxes of the two
Hessian matrices corresponding to the coordinates defined in
tables~\ref{tab:good_coord} and \ref{tab:bad_coord}.

From the values shown, one can conclude that, in the ``properly
defined coordinates'', some convenient characteristics are present: on
one side, the relatively low values of the elements $H_{\chi
\alpha_1}$ and $H_{\chi \alpha_2}$ (and their symmetric ones) indicate
that the soft degree of freedom $\chi$ and the hard ones $\alpha_1$
and $\alpha_2$, which describe the internal structure of the methyl
group, are uncoupled to a reasonable extent; on the other side, the
relatively low value of $H_{\chi \chi}$ compared to $H_{\alpha_1
\alpha_1}$ and $H_{\alpha_2 \alpha_2}$ (a difference of almost an
order of magnitude) proves that $\chi$ may be regarded as soft when
compared to the hard degrees of freedom $\alpha_1$ and $\alpha_2$.

On the contrary, in the ``ill-defined coordinates'', the three
dihedrals are hard, considerably coupled and equivalent.

\section{Conclusions}
\label{sec:conclusions}

Extending the approach of
refs. \cite{PE:Maz1989JBSD,*PE:Aba1989JBSD,*PE:Aba1994JCC} and the
ideas stated in \cite{PE:Cha2002IJQC,PE:Sah2003JMS,PE:Alo2004BOOK}, we
have defined a systematic numeration of the groups, the atoms and the
internal coordinates (termed SASMIC) of organic molecules and,
particularly, of polypeptide chains. The advantages of the rules
herein presented are many-fold:

\begin{itemize}
\item The internal coordinates may be easily cast into conventional
 Z-matrix form and they can be directly implemented into quantum
 chemical packages.
\item The algorithm for numbering allows for automatizing
 and facilitates the coding of computer applications.
\item The modularity of the numeration system in the case of
 polypeptides permits the addition of new residues without
 essentially changing the already numbered items. This
 is convenient if databases of peptide structures need to be
 designed.
\item The set of internal coordinates defined reasonably separate the
 hard and soft movements of organic molecules for arbitrary conformations
 using only topological information.
\end{itemize}

A number of Perl scripts that automatically generate these coordinates
for polypeptide chains are provided as supplementary material. They
may be found at {\tt
http://neptuno.unizar.es/files/public/gen\_sasmic/}.

In addition, we have used the coordinates herein defined and ab initio
Quantum Mechanics to assess the approximation of the free energy
obtained from averaging out the rotameric degree of freedom $\chi$ via
the conventional PES in the protected dipeptide HCO-{\small
L}-Ala-NH$_{2}$. Applying the criterium in ref. \cite{PE:Alo2005JCC},
we have found that approximating $F(\phi,\psi)$ by $V_2(\phi,\psi)$ is
justified up to polypeptides of medium length ($\sim$ 100 residues)
and much computational effort may be saved using the PES instead of
the more realistic free energy. However, the small size of the side
chain of the alanine residue and the fact that the dipeptide is
isolated do not allow to extrapolate this result. For bulkier residues
included in polypeptides, we expect the difference
between $F(\phi,\psi)$ and $V_2(\phi,\psi)$ to be more important.

\vspace{0.2cm} {\small We would like to thank I. Calvo and G.A. Chass,
for illuminating discussions. The numerical calculations have been
performed at the BIFI computing facilities. We thank I. Campos, for
the invaluable CPU time and the efficiency at solving the problems
encountered.}

\vspace{0.2cm} {\small This work has been supported by the Arag\'on
Government (``Biocomputaci\'on y F\'{\i}sica de Sistemas Complejos''
group) and by the research grants MEC (Spain) \mbox{FIS2004-05073} and
MCYT (Spain) \mbox{BFM2003-08532}. P. Echenique is supported by a MEC
(Spain) FPU grant.}

\newpage

\end{document}